\documentclass[12pt]{article}

\begin{document}

\begin{center}

{\Large {\bf II. Calculation of large mass hierarchy \\

\vspace{0,5cm}

from number of extra dimensions}}

\vspace{2cm}

{Boris L. Altshuler}\footnote[1]{E-mail adresses: altshuler@mtu-net.ru \& altshul@lpi.ru}

\vspace{0,5cm}

{\it Theoretical Physics Department, P.N. Lebedev Physical
Institute, \\  53 Leninsky Prospect, Moscow, 119991, Russia}

\vspace{2cm}

{\bf Abstract}
\end{center}

The higher-dimensional generalization of Randall-Sundrum approach with additional positive curvature $n$-dimensional and Ricci-flat $m$-dimensional compuct subspaces is considered in pure gravity theory with metric of space-time and $(p+1)$-form potential as basic fields. Introduction of mass term of $(p+1)$-form potential into the action of co-dimension one brane permits to stabilize brane's position and hence to calculate the value of Planck/electroweek scales ratio. There are no ad hoc too large or too small parameters in the theory; calculated mass hierarchy strongly depends on dimensionalities $m$, $n$ of additional subspaces and its observed large value in 4 dimensions (i.e. for $p=3$) is received in particular in $D13$ ($m=1$, $n=7$) or $D16$ ($m=2$, $n=9$) space-times.

\newpage

\section{Introduction}

\quad In previous Paper I~\cite{IAltsh} the mass term of the antisymmetric $(p+1)$-differential form $A_{p+1}$ was introduced into the brane action in scalar-gravity modification of Randall-Sundrum (RS) model~\cite {Randall} and it was shown that this permits to stabilize brane's position and to calculate the observed large number of Planck/electroweek scales ratio through the moderate value of the dimensionless dilaton-antisymmetric tensor field coupling constant $\alpha$. The theory however is not too predictive since in this model mass hierarchy is quite sensitive to the choice of $\alpha$ (it depends non-analytically on $\alpha$ squared) and physical grounds for $\alpha$ to be of necessary value are unclear. The values of $\alpha$ known from string theory or in $(p+2)$-dimensional scalar-gravity theory received by compactification of a certain number of dimensions in higher dimensional pure gravity theory are far from the value of $\alpha \approx 0,3$ which gives observed mass hierarchy in Paper I.

Situation however improves drastically if higher dimensional space-time contains subspace of non-zero curvature. In this paper we consider $D$-dimensional ($D=(p+1)+n+m+1$) space-time with one extra dimension of the RS theory and two additional Euclidian subspaces: of the constant curvature one ($n$-dimensional sphere $S^{n}$ below) and $m$-dimensional compact Ricci flat one, e.g. $m$-dimensional commutative torus $T^{m}$. For $m=1$ i.e. when $T^{m}=S^{1}$ the observed value of mass hierarchy is received in D13 space-time ($p=3$, $n=7$, $m=1$). And we take the theory without dilaton; the basic fields are metric of $D$-dimensional space-time and $(p+1)$-differential form $A_{p+1}$ (potential of antisymmetric field strength $F_{p+2}$).

The bulk space-time considered in the paper must be limited in the "Randall-Sundrum" direction $z$ ($0<z_{\it min}<z<z_{\it br}$). The most natural way to limit space-time regularly "from below", i.e. at $z=z_{\it min}$, is perhaps to consider the generalized solution with "bolt" at this point (see in the concluding Sec. 5). However, just like in the model considered in \cite{IAltsh}, the choice of $z_{\it min}$ does not influence essentially the calculated value of mass hierarchy. Crucially important for this calculation is the location of the brane limiting space-time "from above" at $z=z_{\it br}$ with $Z_{2}$-symmetry imposed at the brane. For the particular choice of additional subspaces pointed out above this co-dimension one brane has a structure of product of Lorentzian space-time $M^{1,p}$, sphere $S^{n}$ and commutative torus $T^{m}$:

\begin{equation}
\label{1}
M_{\it br}^{(D-1)}=M^{1,p}\,\times\,S^{n}\,\times\,T^{m}.
\end{equation}
Correspondingly there are three different Israel jump conditions - at every brane's subspace each. To make these conditions consistent we are enforced to introduce three terms in the brane action: (1) mass term of the $(p+1)$-form (the introduction of such a term is a novel idea of this paper and of the paper~\cite{IAltsh}); (2) brane's tension term (already used in~\cite{IAltsh} to make Israel conditions consistent); (3) brane's curvature term (non-zero at the $S^{n}$ subspace). These terms must be fine-tuned to each other. It is important to note however that this fine-tuning (see (\ref{24}) below) does not contain too large or too small numbers. Thus dimensional constants characterizing all three terms of brane action are of one and the same scales; in the paper we consider electroweek scale as a fundamental one.

Sec. 2 presents the primary bulk and brane's action, dynamical equations and bulk solution of the model. In Sec. 3 Israel jump conditions and $(p+2)$-form "screening" condition at the brane are written down, mechanism stabilizing brane's position is
demonstrated and brane's location is calculated. In Sec. 4 the Planck/electroweek scales ratio as a function of dimensionalities $m$, $n$ is calculated. The possible generalizations of the model are discussed in Sec. 5.

\section{Description of the model and bulk solution}

\quad Theory used in the paper is described by the action in $D$-dimensional space-time which includes two bulk and three ("hatted") brane terms:

\begin{eqnarray}
\label{2}
S_{(D)}=M^{D-2}\int\Big\{R^{(D)}-\frac{1}{2(p+2)!}F_{p+2}^2 \qquad \mbox{     } \nonumber \\
-\frac{\hat\mu}{2(p+1)!}A_{p+1}^{2}-\hat\sigma+\hat\kappa R^{(D-1)}_{\it br}\Big\}\sqrt{-g^{(D)}}\,d^{D}x+\rm{GH}.
\end{eqnarray}
where "Planck mass" $M$ in $D$ dimensions is
supposed to be of the electroweek scale; $R^{(D)}$ is scalar curvature in $D$ dimensions; $g_{AB}$, $F$, $A$ denote metric, $(p+2)$-form tensor field strength and its potential correspondingly; $R^{(D-1)}_{\it br}$ is scalar curvature of the brane described by the brane's induced metric; $\rm{GH}$ - Gibbons-Hawking surface term. 

$\hat\mu$, $\hat\sigma$ and $\hat\kappa$ in (\ref{2}) are densities located at the brane:

\begin{equation}
\label{3}
\hat\mu=\mu \frac{\delta(z-z_{\it{br}})}{N}, \qquad \hat\sigma=\sigma \frac{\delta(z-z_{\it{br}})}{N}, \qquad \hat\kappa=\kappa \frac{\delta(z-z_{\it {br}})}{N},
\end{equation}
where $N$ is a lapse function of $z$-coordinate transverse to the
brane; dimensional constants $\mu$, $\sigma$, $\kappa$ characterize $(p+1)$-form mass term, brane's tension and strength of brane's gravity correspondingly. 

We take the following anzats for the metric of $D$-dimensional space-time:

\begin{equation}
\label{4}
ds_{(D)}^{2}=b^{2}ds^{2}_{(p+1)}+ a^{2}d\Omega_{(n)}+c^{2}ds^{2}_{(m)}+N^{2}dz^{2},
\end{equation}
where $D=p+n+m+2$, $ds^{2}_{(p+1)}$ is metric of $(p+1)$-dimensional Lorentzian space-time $M^{1,p}$, $d\Omega_{(n)}$ is metric of unit $n$-dimensional sphere and $ds^{2}_{(m)}$ is metric of $m$-dimensional compact Euclidian Ricci-flat manifold.
The antisymmetric $(p+1)$-form potential possesses non-zero component on $M^{1,p}: $

\begin{equation}
\label{5}
A_{p+1}=f(z)\epsilon_{\mu_{1}\cdots\mu_{p+1}},
\end{equation}
wherefrom

\begin{equation}
\label{6}
F_{p+2}=f'(z)\epsilon_{\mu_{1}
\cdots\mu_{p+1}z}=Q\,\frac{b^{p+1}N}{a^{n}c^{m}}\,\epsilon_{\mu_{1}\cdots\mu_{p+1}z},
\end{equation}
prime means derivation over $z$, the last equality in (\ref{6}) is the solution of bulk "Maxwell" equation (\ref{14}) below, "charge" $Q$ is an arbitrary constant of the solution. From (\ref{5}), (\ref{6}) with account of (\ref{4}) and negative signature of the Lorentzian manifold $M^{1,p}$ it follows:

\begin{equation}
\label{7}
\frac{A_{p+1}^2}{(p+1)!}=-\frac{f^2}{b^{2p+2}},
\end{equation}

\begin{equation}
\label{8}
\frac{F_{p+2}^2}{(p+2)!}=-\frac{f'^{2}}{b^{2p+2}N^{2}}=
-\frac{Q^{2}}{a^{2n}c^{2m}}.
\end{equation}

From three brane's subspaces in (\ref{1}) only sphere $S^{n}$ gives contribution to the brane's curvarture. Hence:

\begin{equation}
\label{9}
R^{(D-1)}_{\it br}=\frac{n(n-1)}{a^{2}}.
\end{equation}

After these preliminaries we shall write down dynamical equations following from the action (\ref{2}). Those equations are: constraint (\ref{10}), three second order Einstein equations (\ref{11})-(\ref{13}) for metric's scale functions $b(z)$, $a(z)$, $c(z)$ and equation (\ref{14}) for potential $A_{p+1}$ defined in (\ref{5}). Expressions (\ref{7}), (\ref{8}), (\ref{9}) for $A^{2}$, $F^{2}$ and brane's curvature are taken into account in writing down Eqs. (\ref{10})-(\ref{14}), we remind that  $D-2=p+n+m$, prime means derivation over $z$:

\begin{eqnarray}
\label{10}
&&-\frac{n(n-1)}{2a^{2}}+\frac{1}{2N^{2}}\Bigg[p(p+1)\frac{b'^2}{b^2}+n(n-1)\frac{a'^2}{a^2}+m(m-1)\frac{c'^2}{c^2} \nonumber \\
&&+2n(p+1)\frac{b'a'}{ba}+2m(p+1)\frac{b'c'}{bc}+2mn\frac{a'c'}{ac}\Bigg]=-\frac{Q^2}{4a^{2n}c^{2m}};
\end{eqnarray}
\\
\begin{eqnarray}
\label{11}
&&-\frac{1}{N^2}\Bigg[\frac{b''}{b}+\frac{b'}{b}\Bigg(-\frac{N'}{N}+p\frac{b'}{b}+n\frac{a'}{a}+m\frac{c'}{c}\Bigg)\Bigg]= \nonumber \\
&&=\frac{1}{2(D-2)}\,\Bigg[-\frac{Q^2}{a^{2n}c^{2m}}\,(n+m-1) \nonumber \\
&&-\hat\mu\,\frac{f^2}{2b^{2p+2}}\,(2n+2m-1)+\hat\sigma+\hat\kappa\,\frac{n(n-1)}{a^2}\Bigg];
\end{eqnarray}
\\
\begin{eqnarray}
\label{12}
&&\frac{n-1}{a^2}-\frac{1}{N^2}\Bigg[\frac{a''}{a}+\frac{a'}{a}\Bigg(-\frac{N'}{N}+(p+1)\frac{b'}{b}+(n-1)\frac{a'}{a}+m\frac{c'}{c}\Bigg)\Bigg]= \nonumber
\\
&&=\frac{1}{2(D-2)}\,\Bigg[\frac{Q^2}{a^{2n}c^{2m}}\,(p+1) \nonumber
\\
&&+\hat\mu \frac{f^2}{2b^{2p+2}}\,(2p+1)+\hat\sigma-\hat\kappa\,\frac{n(n-1)}{a^2}\,\frac{2p+2m+n}{n}\Bigg];
\end{eqnarray}
\\
\begin{eqnarray}
\label{13}
&&-\frac{1}{N^2}\Bigg[\frac{c''}{c}+\frac{c'}{c}\Bigg(-\frac{N'}{N}+(p+1)\frac{b'}{b}+n\frac{a'}{a}+(m-1)\frac{c'}{c}\Bigg)\Bigg]= \nonumber
\\
&&=\frac{1}{2(D-2)}\,\Bigg[\frac{Q^2}{a^{2n}c^{2m}}(p+1) \nonumber
\\
&&+\hat\mu\,\frac{f^2}{2b^{2p+2}}(2p+1)+\hat\sigma+\hat\kappa\,\frac{n(n-1)}{a^2}\Bigg];
\end{eqnarray}
\\
\begin{equation}
\label{14}
\frac{1}{J}\Bigg[\frac{Jf'}{b^{2p+2}N^{2}}\Bigg]'=\hat \mu\, \frac{f}{b^{2p+2}}, \qquad J\equiv b^{p+1}a^{n}c^{m}N.
\end{equation}
\\
The bulk solution of these equations is given below for the choice of $z$ to be a proper distance i.e. for the choice $N=1$ in the metric (\ref{4}):

\begin{equation}
\label{15}
b=\left(\frac{z}{l}\right)^{u}, \quad a=\frac{m}{n+m-1}\,z, \quad c=\left(\frac{z}{l}\right)^{-(n-1)/m},
\end{equation}
\\
\begin{equation}
\label{16}
f=\left[\frac{2(p+n+m)}{(n+m-1)(p+1)}\right]^{1/2}\left(\frac{z}{l}\right)^{(p+1)u}+ const,
\end{equation}
\\
where
\begin{equation}
\label{17}
u \equiv \frac{(n+m-1)(n-1)}{(p+1)m}
\end{equation}
and length $l$ is an arbitrary constant of the solution determined by "charge" $Q$ of the $(p+2)$-form by the relation:

\begin{equation}
\label{18}
\frac{Q^2}{l^{2(n-1)}}=\frac{2m^{2n-2}(n-1)^{2}(p+n+m)}{(n+m-1)^{2n-1}(p+1)}.
\end{equation}

The important feature of this solution is the essential anisotropy in dependence on $z$ of scale factors of different subspaces of metric (\ref{4}). In case $m=0$ solution (\ref{15}), (\ref{16}) must come to the well known supergravity solution where $a=const$, $F_{p+2}^{2}=const$ and $b$ depends on $z$ in the AdS-type exponential way. This limit is not immediately seen in (\ref{15}), (\ref{16}). However it becomes transparent if this solution is rewritten in the lapse-gauge $N=b^{-1}$; the generalized version of bulk solution in this gauge is written down in Sec. 5. Space-time described by the metric (\ref{15}) posseses singularity at $z=0$, hence it must be limited at some $z=z_{\it min}>0$ - see discussion in Sec. 5.   

\section{Jump conditions and stabilization of brane's position}

\quad Now we shall consider the role of "hatted" $\delta$-function brane terms in the RHS of equations (\ref{11})-(\ref{14}). Integration of these equations over brane's position with account of definitions (\ref{3}) and $Z_{2}$-symmetry imposed at the brane gives immediately three Israel jump conditions (\ref{19})-(\ref{21}) for the first derivatives of scale factors $b, a, c$ of three brane's subspaces and "screening" condition (\ref{22}) for $(p+2)$-form (cf. (\ref{5}), (\ref{6})) which is a crucial one for brane's stabilization:

\begin{equation}
\label{19}
\frac{2}{N^{2}} \frac{b'}{b}=\frac{1}{2(D-2)}\,\Bigg[-\frac{\mu}{N}\frac{f^2}{2b^{2p+2}}(2n+2m-1)+\frac{\sigma}{N}+\frac{\kappa}{N}\frac{n(n-1)}{a^{2}}\Bigg],
\end{equation}

\begin{equation}
\label{20}
\frac{2}{N^{2}} \frac{a'}{a}=\frac{1}{2(D-2)}\,\Bigg[\frac{\mu}{N}\frac{f^2}{2b^{2p+2}}(2p+1)+\frac{\sigma}{N}-\frac{\kappa}{N}\frac{n(n-1)}{a^{2}}\,\frac{2p+2m+n}{n}\Bigg],
\end{equation}

\begin{equation}
\label{21}
\frac{2}{N^{2}} \frac{c'}{c}=\frac{1}{2(D-2)}\,\Bigg[\frac{\mu}{N}\frac{f^2}{2b^{2p+2}}(2p+1)+\frac{\sigma}{N}+\frac{\kappa}{N}\frac{n(n-1)}{a^{2}}\Bigg],
\end{equation}

\begin{equation}
\label{22}
-\frac{2}{N^{2}}f'=\frac{\mu}{N}f.
\end{equation}
\\
Eqs. (\ref{19})-(\ref{22}) are valid at the brane's position $z_{\it br}$ which is determined from (\ref{22}) and (\ref{16}) through dimensional constant $\mu$ and dimensionalities $m, n$:

\begin{equation}
\label{23}
z_{\it br}=-\frac{2(n+m-1)(n-1)}{m\mu}.
\end{equation}

It is seen that we must put $\mu<0$ in the action (\ref{2}). In determining $z_{\it br}$ we put equal to zero the arbitrary $const$ in the solution for $f$ in (\ref{16}). In some future investigations this $const$ may be considered non-zero (e.g. to make potential $A_{p+1}$ equal to zero at $z=z_{\it min}$ i.e. at the "bolt" of generalized solution written down in Sec. 5). In any case because of growth of $f$ with $z$ in (\ref {16}) this will give just a small corrections to the brane's position (\ref {23}) and hence to the calculated value of the Planck/electroweek scales ratio.

Substitution of the expression (\ref{23}) for $z_{\it br}$ and solution (\ref{16}) (with $const=0$ in this expression for $f$) into Israel jump conditions (\ref{19})-(\ref{21}) shows that only two of them are independent and give the following consistency fine-tuning conditions for brane action parameters $\mu$, $\sigma$, $\kappa$ in (\ref{2}):

\begin{equation}
\label{24}
\kappa\mu=4, \qquad \frac{\sigma}{\mu}=-\frac{n}{n-1}+\frac{(p+n+m)}{(p+1)(n+m-1)}, \nonumber \\
\end{equation}
where $\mu<0$, $\sigma>0$, $\kappa<0$.

\section{Calculation of mass hierarchy in 4 dimensions}

\quad To calculate from the action (\ref{2}) the Planck/electroweek scales ratio (further on in this Section we shall put $(p+1)=4$) we must integrate over $z$ the "4-dimensional" term of the curvature $R^{(D)}$ ($D=5+n+m$) in the action (\ref{2}) which is equal to $\tilde{R}^{(4)}/b^2$, where $\tilde{R}^{(4)}$ is a scalar curvature in 4-dimensional space-time. Using metric (\ref{4}) specified in (\ref{15}) we get from (\ref{2}):

\begin{eqnarray}
\label{25}
&&M_{\rm Pl}^{2}=M^{3+n+m}V_{m}\Omega_{n}\int_{z_{\it{min}}}^{z_{\it{br}}}b^{2}a^{n}c^{m}\,dz= \nonumber
\\
&&=M^{3+n+m}l^{n}V_{m}\Omega_{n}\Bigg(\frac{m}{n+m-1}\Bigg)^{n}\,\int_{z_{\it{min}}}^{z_{\it{br}}}\left(\frac{z}{l}\right)^{2u+1}\,dz,
\end{eqnarray}
\\
where $u$ is determined in (\ref{17}), $l$ - in (\ref{18}), $V_{m}$ is dimensional volume of compact Ricci-flat subspace, $\Omega_{n}$ - volume of $n$-dimensional unit sphere. The choice of the lower limit $z_{\it{min}}$ does not effect essentially the value of integral in (\ref{25}) and we shall put $z_{\it{min}}=0$ below. Upper limit $z_{\it{br}}$ is determined in (\ref{23}). With this we receive finally from (\ref{25}) the Planck/electroweek scales ratio in 4 dimensions as a function of dimensionalities $m$, $n$ and dimensional constants of the theory $M$, $\mu$, $l$, $V_{m}$:

\begin{eqnarray}
\label{26}
&&\frac{M_{\rm{Pl}}}{M}=\frac{[(Ml)^{(n+1)}M^{m}V_{m}\Omega_{n}]^{1/2}}{(-\mu \,l)^{\delta}(2\delta)^{1/2}}\Bigg(\frac{m}{n+m-1}\Bigg)^{n/2}\,\Bigg[\frac{2(n+m-1)(n-1)}{m}\Bigg]^{\delta}, \nonumber
\\
&& \qquad  \qquad  \qquad  \delta \equiv 1+\frac{(n+m-1)(n-1)}{4m}.
\end{eqnarray}
\\
Following the general approach of having all arbitrary "input" dimensional constants of the primary action and of the solution to be of one and the same order and with a goal to present a quantitative estimation of the result (\ref{26}) we put in (\ref{26})

\begin{equation}
\label{27}
-\mu=M, \qquad l=M^{-1}, \qquad V_{m}\Omega_{n}=M^{-m},
\end{equation}
\\
which gives in particular for two cases of minimal values of $m=1$ and $m=2$ and for certain values of $n$ in each case:
\\
\\
1) $m=1$:

\begin{equation}
\label{28}
\frac{M_{\rm{Pl}}}{M}=1,5 \cdot 10^{12}\,(n=6); \quad 3,1 \cdot 10^{18}\,(n=7); \quad 2,4 \cdot 10^{26}\, (n=8).
\end{equation}
\\
2) $m=2$:

\begin{equation}
\label{29}
\frac{M_{\rm{Pl}}}{M}=5,4 \cdot 10^{12}\,(n=8); \quad 1,3 \cdot 10^{17}\,(n=9); \quad 1,9 \cdot 10^{22}\, (n=10).
\end{equation}
\\
Very strong dependence of mass hierarchy on dimensionalities of subspaces is perhaps the most interesting feature of the proposed approach. As it is seen from (\ref{28}), (\ref{29}) the value of mass hierarchy close to the observed one is received in particular in $D13$ ($m=1$, $n=7$) or in $D16$ ($m=2$, $n=9$) space-times.

\section{Discussion}

\quad One of the natural ways to generalize the bulk metric (\ref{4}), (\ref{15}) is to introduce the additional compact space-like direction with a role of Euclidian "time" like it was done e.g. in~\cite{Louko},~\cite{Burgess}. In Sec. 5 of Paper I~\cite{IAltsh} this sort of solution was written down for scalar-gravity theory, here we shall do it for space-time with additional subspaces in a theory without dilaton considered above. Thus let us view metric $ds^{2}_{(p+1)}$ of the Lorentzian subspace of the space-time (\ref{4}) as a product of $p$-dimensional Lorentzian space-time $M^{1,(p-1)}$ and a circle $S^{1}$ with coordinate $y$. The generalization of bulk solution (\ref{15}) (written down in "Schwarzschild" lapse-gauge) is described by the metric:

\begin{equation}
\label{30}
ds_{(D)}^{2}=b^{2}ds^{2}_{(p)}+\Delta dy^{2}+\frac{dr^2}{\Delta}+ a^{2}d\Omega_{(n)}+c^{2}ds^{2}_{(m)},
\end{equation}
where
\begin{equation}
\label{31}
b \sim r^{\xi}, \quad a \sim r^{1-\xi}, \quad c \sim r^{-(1-\xi)(n-1)/m}, \quad \Delta (r)=C_{1}r^{2\xi}+ C_{2}r^{(1-p)\xi}.
\end{equation}
In (\ref{31}) $\xi=u/(1+u)$ and $u$ is a function of dimensionalities $p$, $n$, $m$ determined in (\ref{17}). 

Term "$C_{1}$" in $\Delta$ in (\ref{31}) is generated by the bulk potential $\sim F_{p+2}^{2}$; in case it is the only term in $\Delta$ Eqs. (\ref{30}), (\ref{31}) give the bulk solution (\ref{15}) rewritten in the lapse-gauge $N=b^{-1}$. It is easily seen from (\ref{17}) that for $m=0$ we have $\xi \equiv u/(1+u)=1$ and metric (\ref{30}), (\ref{31}) comes to the well known Schwarzschild-AdS solution:

\begin{equation}
\label{32}
ds_{(D)}^{2}=b^{2}ds^{2}_{(p)}+\Delta dy^{2}+\frac{dr^2}{\Delta}+ a^{2}d\Omega_{(n)},
\end{equation}
where
\begin{equation}
\label{33}
b \sim r, \quad a=const, \quad \Delta=C_{1}r^{2}+C_{2}r^{1-p}.
\end{equation}

Term "$C_{2}$" in $\Delta$ in (\ref{31}) is a generalization of conventional Schwarzschild term. With a choice of arbitrary constant $C_{2}$ it is always possible to make
$\Delta(r)=0$ at some $r=r_{0}$ called "bolt" which, as it was said above, may be used to limit space-time regularly "from below" in the "Randall-Sundrum" direction $r$ (also named $z$ in the bulk of the paper).

There are only two terms in $\Delta$ in (\ref{31}) or (\ref{33}) contrary to the solution given e.g. in~\cite{Louko} where there are four terms in $\Delta$ - one additional term is generated by the Maxwell field and another one (which is  a constant) is generated by non-zero cosmological term of the Lorentzian $M^{1, (p-1)}$ space-time. In our case however because of presence of additional subspaces of the $D$-dimensional space-time the corresponding "Maxwell field" generalization of metrics (\ref{30}), (\ref{31}) or (\ref{32}), (\ref{33}) is not immediately seen. The non-zero cosmological term of the Lorentzian space-time $M^{1,(p-1)}$ will result in the well known additional constant term in $\Delta$ in the conventional solution (\ref{32}), (\ref{33}) (case $m=0$). However when there is additional subspace of dimensionality $m>0$ we did not manage to find generalization of the solution (\ref{30}), (\ref{31}) to the case of non-zero curvature of the Lorentzian subspace $M^{1,(p-1)}$. To find such a solution would be especially interesting in view of the discussed in~\cite{Altshuler} possibility to receive the small observed value of the positive cosmological constant from adjustment of Israel conditions in the "brane-bolt" model.

And of course it would be very interesting if some physical grounds were pointed out for appearance in the brane action of the antisymmetric tensor field mass term used in~\cite{IAltsh} and in the present paper.

\section*{Acknowledgements} Athour is greatful to Vladimir Nechitailo for assistance. This work was partially supported by the grant LSS-1578.2003.2

\end{document}